\documentclass[runningheads]{llncs}
\usepackage{eccv}

\usepackage{eccvabbrv}

\usepackage{graphicx}
\usepackage{booktabs}
\usepackage{pdfpages}

\usepackage[hidelinks]{hyperref}

\usepackage{orcidlink}

\usepackage{verbatim}
\usepackage{multirow}
\usepackage[normalem]{ulem}
\usepackage[table,xcdraw]{xcolor}
\usepackage{diagbox}
\usepackage{amsmath}
\usepackage{amssymb}
\usepackage[]{hyperref}
\usepackage{array}
\usepackage[T1]{fontenc}
\usepackage{booktabs}
\usepackage{etoolbox}

\usepackage{changepage}
\let\oldabstract\abstract
\let\oldendabstract\endabstract
\renewenvironment{abstract}
  {\begin{adjustwidth}{-0.1cm}{-0.1cm}\oldabstract}
  {\oldendabstract\end{adjustwidth}}
  
\begin{document}

\title{Rendering Novel Views of MRI Using 3D Gaussian Splatting} 

% \titlerunning{Rendering Novel Views of MRI Using 3D Gaussian Splatting}

\author{Robin Y. Park\inst{\,*,1,\raisebox{-0.25ex}{\href{mailto:robinpark@robots.ox.ac.uk}{\includegraphics[width=0.7em]{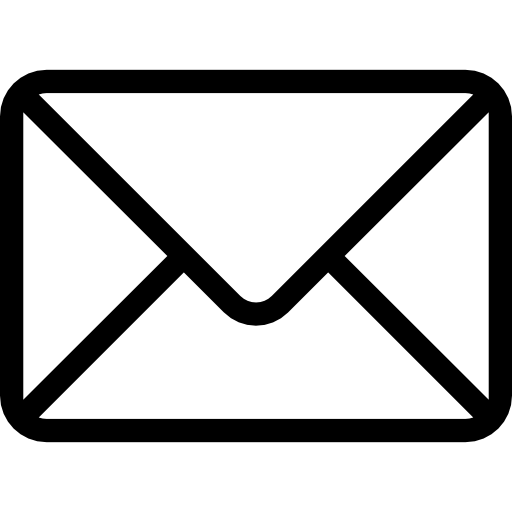}}}} \and
Mark C. Eid\inst{\,*,1,2,\,\raisebox{-0.25ex}{\href{mailto:markeid@robots.ox.ac.uk}{\includegraphics[width=0.7em]{mail.png}}}} \and \\[2pt] Rhydian Windsor\inst{\,1} \and Amir Jamaludin\inst{\,1} \and Ana I.L. Namburete\inst{\,2} \and \\ João~F.~Henriques\inst{\,1} \and Andrew Zisserman\inst{\,1}}

\renewcommand{\thefootnote}{\fnsymbol{footnote}}
\setcounter{footnote}{1} 
\footnotetext{These authors contributed equally.}
\renewcommand{\thefootnote}{\arabic{footnote}}

\authorrunning{R.~Park, M. Eid et al.}

\institute{Visual Geometry Group, University of Oxford \and
Oxford Machine Learning in NeuroImaging Lab, University of Oxford\\
% \email{\{robinpark,markeid\}@robots.ox.ac.uk}}
$^{\raisebox{-0.25ex}{\includegraphics[width=0.7em]{mail.png}}}$\{\href{mailto:robinpark@robots.ox.ac.uk}{robinpark},\href{mailto:markeid@robots.ox.ac.uk}{\,markeid}\}@robots.ox.ac.uk
}

\maketitle

{\centering\textit{AI4M3D Spotlight @ ECCV 2026}\par}
\mbox{}\par

\begin{abstract}
  The objective of this paper is to improve radiological gradings measured on MRIs of spines, by resampling scans so that the new view planes are better aligned with the target anatomy than the original sparse images. To this end, we adapt 3D Gaussian Splatting to form a volumetric reconstruction starting from non-aligned MRIs to render imaging planes aligned with the anatomy relevant for clinical evaluation. The novel view plane is optimal for diagnostic radiological grading of the target anatomy, whereas the original MRI is not. The resampled scans are then used to predict ordinal severity grades of localised stenosis conditions in spinal MRIs. We compare our method against (1) Voxel Interpolation, which takes the average of inverse-distance weighted nearest neighbour intensities for each target coordinate, and (2) Cubic B-spline resampling, which estimates target points using a smooth piecewise-cubic function built from the source points. Experiments show that across all stenosis conditions, resampled scans using Gaussian Splatting produce more accurate stenosis gradings compared to the raw scans which do not include the complete anatomy in-plane, as well as images resampled using Voxel Interpolation or Cubic B-spline resampling. 
  \keywords{MRI reconstruction \and Gaussian splatting \and Stenosis grading}
\end{abstract}

\section{Introduction}
\label{sec:intro}

Lumbar spinal stenosis is the narrowing of the spinal canal in the lower back, affecting more than 100 million patients worldwide and a leading cause of disability in older adults~\cite{katz2022diagnosis, sobanski2023presentation}. More severe cases often require magnetic resonance imaging (MRI) to assess the extent of disease and guide treatment decisions~\cite{skasiadek2024degenerative}. The severity of disease is usually graded on an ordinal, categorical scale. For example, lumbar foraminal stenosis is typically graded on a four-point scale, consisting of absent (normal), mild, moderate and severe~\cite{lee2010practical}. Accurate radiological grading depends on image quality with the stipulation that the anatomy of interest is clearly visible in frame~\cite{kastryulin2023image}. 

While 2D MRI slices are made up of pixels, each pixel corresponds to a voxel of visualised anatomy, representing a 3D volume of length, width and slice thickness. In diagnostic MRI, the voxels represented are often anisotropic, meaning that the distance between 2D slices is typically much larger than in-plane voxel length and width~\cite{jia2017new}. As a result, reconstruction using MRI acquired in standard clinical practice requires more consideration of voxel dimensions and slice spacing, compared to scanning modalities typically acquired with isotropic resolution such as CT.  

\begin{figure}[h]
\centering 
\includegraphics[width=0.9\textwidth]{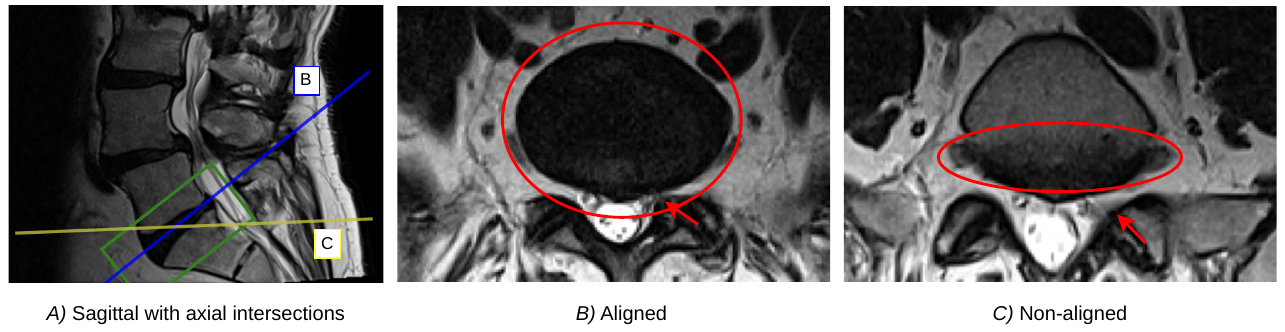}
\caption{\label{fig:ex-paired} \textbf{Paired axial scans:} Slices from two T2w axial scans of the same patient, with moderate left subarticular stenosis at the L5-S1 intervertebral disc (IVD) level (see Fig.~\ref{fig:stenosis-regions} for a description of spinal anatomy and stenosis zones). The aligned scan \textit{B} shows the full disc (red ellipse) while the non-aligned scan \textit{C} only depicts a part of the disc, obscuring the presence of stenosis (red arrow in \textit{B}). The grey structure above the disc fragment in \textit{C} is the vertebra. The blue line on the T2w sagittal scan \textit{A} indicates the position of an aligned slice (\textit{B}), and the yellow line corresponds to a non-aligned slice (parallel to \textit{C}). The green box in \textit{A} depicts the region of interest (ROI).} 
\end{figure}

Axial scans are often obtained as a single parallel acquisition, even for the lumbar spine where the natural curvature means that lower discs are not aligned with levels above~\cite{van2005observer,landauer2022diagnostic}. This means that crucial anatomical details necessary for accurate grading may go missed when the imaging planes are misaligned with the target anatomy~\cite{sarma2025aoi}. However, acquiring anatomically aligned MRI first requires a scan through the region of interest (ROI) to determine the correct position, so takes longer than a single whole spine acquisition~\cite{park2013practical, park2015diagnostic}. Fig.~\ref{fig:ex-paired} shows aligned and non-aligned axial scans, and how they intersect the ROI. 

This paper explores the application of 3D Gaussian Splatting for resampling axial MRI at angles that are aligned with the disc, for better radiological grading. Our objective is to produce aligned scans (Fig.~\ref{fig:ex-paired}B) starting from non-aligned scans of the same patient (Fig.~\ref{fig:ex-paired}C), and then use the reconstructed aligned scans (which we term ``Resampled'' images) to grade four different types of localised stenosis.

Our method is compared against two baselines for image quality and radiological grading performance: (1) Voxel Interpolation, which uses the mean of inverse-distance weighted nearest neighbour intensities per target coordinate, and (2) Cubic B-Spline resampling, which computes target intensities as a smooth, cubic-weighted estimate of neighbouring source intensities. We evaluate the resampled images using (1) the image quality with the Learned Perceptual Image Patch Similarity (LPIPS) metric; and (2) their performance on four downstream spinal stenosis grading. Our results show that 3D Gaussian Splatting resampled images achieve better localised stenosis grading performance than non-aligned scans that do not fully capture the region of interest in-plane.

\subsection{Related work}
\subsubsection{Gaussian Splatting and non-projection-based image reconstruction.}
Gaussian Splatting was originally developed for {\em projective imaging} of scenes~\cite{kerbl20233d}, and quickly replaced implicit neural radiance fields (NeRFs~\cite{mildenhall2020nerf}) with explicit 3D Gaussians and fully differentiable rasterisation, enabling fast, high-quality rendering. Since then there have been a multitude of improvements~\cite{bao3DGaussianSplatting2025,fei3DGaussianSplatting2025}, and in a similar manner to NeRFs~\cite{iddrisuNerfMRI,eid_rapidvol_2025}, Gaussian Splatting was applied to the medical domain for 3D reconstruction of endoscopic videos~\cite{guo2024free, huang2025surgtpgs, bonillaGaussianPancakes2024}, CT and X-Ray scans~\cite{zha_r2-gaussian_2024, leonardis_radiative_2025}, and anatomical surface reconstructions using multimodal scans~\cite{marzol2025medgs}. Liu \etal~\cite{liu_survey_2025} provide a more detailed survey. Most relevantly, it was recently adapted for non-projection-based images for the reconstruction of ultrasound~\cite{eid2025ultragauss}, which like MRI (and unlike X-Ray and endoscopy) is a {\em non-projection}-based modality. 

This work is the first to apply 3D Gaussian Splatting to spinal MRI reconstruction, another non-projection-based modality. MRI does not form images by tracing light rays and accumulating them through a pin-hole projection camera, but relies on the resonance of hydrogen atoms in 3D space, and this complex image formation model is often best approximated as a greyscale voxel volume. This makes many projection-based prior works unsuitable. MRI reconstructions also require significantly more contrast and detailing as compared to ultrasound~\cite{devriesImagingPrematureBrain2013, blondiauxFetalCerebralImaging2013, tawfikDiagnosticValueSpinal2020}. Additionally, the average spacing between input images used to form the MRI reconstructions is much larger than that used by Eid \etal~in UltraGauss~\cite{eid2025ultragauss} ($\sim4$mm vs. 0.6 mm). The MRI voxel dimensions are also variable from scan to scan, unlike with the ultrasound data of~\cite{eid2025ultragauss}, meaning our reconstruction model must be able to generalise better. This all makes the transition from ultrasound volumetric reconstruction to MRI non-trivial.

\subsubsection{Radiological stenosis grading.}
The need for rapid MRI evaluation has driven development of automated radiological stenosis grading, using convolutional ~\cite{jamaludin2017spinenet,lu2018deep} and transformer-based networks~\cite{windsor2024automated, windsor2022context, park2025multi}. 

SpineNet~\cite{windsor2024automated, jamaludin2017spinenet} provides automated radiological gradings including stenosis, but only on T2-weighted (T2w) sagittal MRI. DeepSPINE~\cite{lu2018deep} can produce stenosis grades for T2w axial scans, but uses 3D convolutions for volumetric feature extraction, which cannot be applied to 2D images and assumes the thickness of the cut to be constant across images (contrary to MRI samples used in clinical practice). A more detailed survey of automated stenosis grading methods can be found in~\cite{verheijen2025artificial}. We use the transformer-based multi-view grading model proposed by Park \etal~\cite{park2025multi}, and apply it to Axial T2-weighted (T2w) MRI. Rather than solely maximising grading accuracy, we analyse how model performance varies with how well the input images show the anatomy of interest. 

\section{Dataset}
\begin{figure}[h]
\centering 
\includegraphics[width=0.9\textwidth]{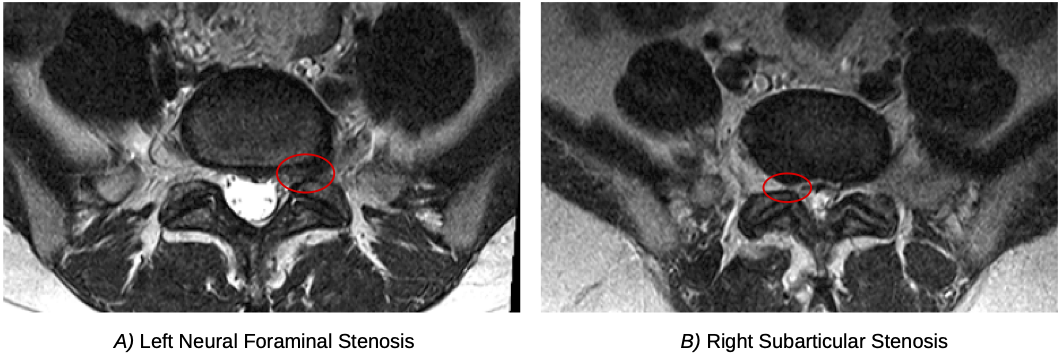} 
\caption{\label{fig:ex-stenosis} \textbf{Examples of stenosis:} Axial slices showing severe stenosis at L5-S1. See Fig.~\ref{fig:stenosis-regions} for a description of anatomical structures and zones where different localised stenosis conditions occur. Notice in \textit{A}, there is extrusion of the disc into the foraminal region (dark oval in the center of the image), causing diminished space compared to the corresponding region on the other side of the spinal canal, indicating severe left neural foraminal stenosis (marked with a red ellipse). \textit{B} shows extrusion and reduced space in the subarticular zone compared to the corresponding region on the other side of the canal, indicating severe right subarticular stenosis (also marked by a red ellipse). Axial MRI is conventionally interpreted from the inferior-to-superior perspective.} 
\end{figure}

We use a publicly available dataset of anonymised MR scans provided by the Radiological Society of North America (RSNA)~\cite{richards2026rsna}, comprising of 1,975 patients (6,294 T1w, T2w and STIR sagittal and axial scans) from 12 institutions, with vertebral-level grading labels contributed by over 50 expert clinicians from 15 different countries. Annotations cover L1-L2 to L5-S1 for five conditions: spinal canal stenosis, left/right neural foraminal stenosis, and left/right subarticular stenosis. The dataset was provided and annotated as part of the Lumbar Spine Degenerative Classification AI Challenge (2024), to assess the presence and severity of stenosis conditions that contribute to low back pain, a leading cause of disability worldwide~\cite{katz2022diagnosis, lin2006disability}.  We focus on the last lumbar disc, L5-S1, which is often the most misaligned with the Z-axis and the site of 75\% of all neural foraminal stenosis~\cite{takahashi2022foraminal}. Due to low occurrence at our level of interest (L5-S1), spinal canal stenosis is excluded from our experiments. Each condition is graded as normal/mild (single class), moderate, or severe. Fig.~\ref{fig:ex-stenosis} shows examples of severe neural foraminal stenosis and subarticular stenosis. 

We focus on the axial scans in this dataset due to a unique property: some patients have multiple axial scans from different views within the same scan session (see Fig.~\ref{fig:3d-planes}). Where the image acquisition plane is adjusted to be \textbf{aligned} to each disc, the full disc is visible in-slice (see Fig.~\ref{fig:ex-paired}B). Where the image acquisition plane is uniform through the spine, upper lumbar discs are fully visible, but lower discs like L5-S1 that are ``tilted'' following the spine's natural curvature are \textbf{non-aligned} and only partially captured in-slice (see Fig.~\ref{fig:ex-paired}C).  While most patients in this dataset have an aligned scan, this is not the case in many spinal research datasets, leading to the exclusion of L5-S1 in training deep learning models~\cite{kim2025translation, hong2022lumbar}.

Another key feature is that, as in most diagnostic MRI, the voxels are anisotropic~\cite{van2012super}. Voxels are densely spaced in-slice but much more sparse perpendicular to each plane. Voxel dimensions also vary widely across scans, likely as the data were collected from multiple institutions with diverse imaging protocols.

With paired samples, we can compare a grading model's performance on \textbf{aligned (A)}, \textbf{non-aligned (N)} and \textbf{resampled (R)} images for the same patient to assess the effectiveness of resampling for downstream stenosis grading.

\section{Reconstruction of MR volumes using Gaussian Splatting}

\begin{figure}[!h]
\centering
\includegraphics[width=0.9\textwidth]{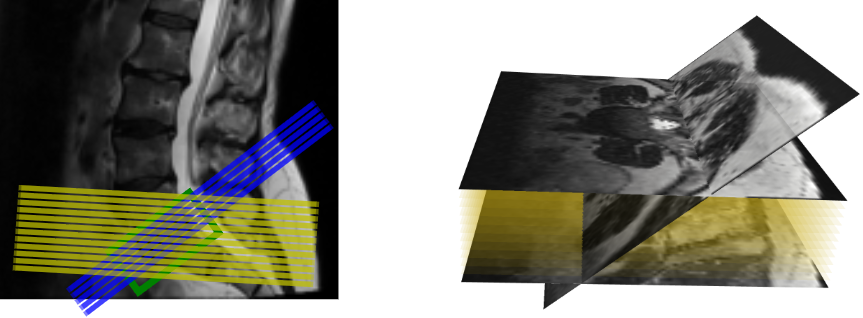}
\caption{\label{fig:3d-planes} \textbf{Axial scans intersecting in 2D \& 3D:} The oblique plane in the 3D plot (right image) shows the mid-slice of the axial volume aligned with the disc (blue lines in left image). The yellow planes are non-aligned slices of a volume that does not follow the angle of the disc, but intersect the intervertebral disc region of interest (green box in left image).} 
\end{figure}

\noindent In this section, we describe how we apply 3D Gaussian Splatting to reconstruct MRI from novel views using sparse source volumes. The reconstruction objective is to use non-aligned axial volumes to emulate scans taken from the aligned view. The ``target'' planes to resample are shown by blue lines in Fig.~\ref{fig:3d-planes}, while ``source'' planes are the non-aligned slices (yellow in Fig.~\ref{fig:3d-planes}). Target voxels are defined by the novel-view plane's 3D position, with voxel intensities sampled from the reconstructed non-aligned source volume. With both the target and source volumes, the 3D image-plane poses are defined relative to the mid-slice of each acquisition, taking into account the anisotropic nature of the voxels (slice spacing is used to define the planes in space). 

To form a 3D reconstruction and sample 2D views from it, we aggregate 3D Gaussians (each with a learnable 3D mean $\mu_{i}$, 3D covariance $\Sigma_{i}$, greyscale colour $c_i$, and opacity $\alpha_i$), onto each pixel of a cross-sectional image plane. Specifically, the source slices are first fed in with their 3D poses. These poses, as with the Gaussians' means and covariances, are defined relative to the mid-slice of the acquisition. For each pose, the Gaussians' parameters are then transformed from the mid-slice frame of reference to each image plane's frame of reference ($\mu_{i}, \Sigma_{i} \to \mu_{i}', \Sigma_{i}'$). Each pixel on the image plane with 3D position $x'$ then has its colour $c_{\mathrm{Pixel}}$ computed by aggregating the surrounding Gaussians within a 95\% probability distribution as per Eqs.~\ref{eq:mahalanobis-3d} - \ref{eq:colour-combine}: 

\begin{equation}
\hat{\alpha}_{i}\left(x'\right)=\alpha_{i}\exp\left(-\frac{1}{2}\right.\underbrace{\left(x'-\mu_{i}'\right)^{T}\left(\Sigma_{i}'\right)^{-1}\left(x'-\mu_{i}'\right)}_{\textrm{3D squared Mahalanobis distance}}\left.\vphantom{\frac{1}{2}}\right)\label{eq:mahalanobis-3d}
\end{equation}
\begin{equation}
\hat{\alpha}\left(x'\right) =\sum_{i}^{N}\hat{\alpha}_{i}\left(x'\right)+\alpha_{\mathrm{BG}} \,\text{;}\quad
c_{\mathrm{Pixel}}(x')=\frac{1}{\hat{\alpha}\left(x'\right)}\left(\sum_{i}^{N}\hat{\alpha}_{i}\left(x'\right)c_{i}+\alpha_{\mathrm{BG}}c_{\mathrm{BG}}\right)
\label{eq:colour-combine}
\end{equation}

\noindent where a small background alpha term $\alpha_{BG}$ is added to prevent division by zero if any areas lack Gaussians, and the background colour is chosen to be black (i.e. $c_{BG}=0$). $N$ is the number of Gaussians which contributed to pixel $x'$s colour.

Notably, we keep the 3D Gaussians as they are (\ie in 3D not 2D), and work with the 3D Mahalanobis distance between the voxel centre and Gaussian mean to calculate that Gaussian's contribution to the final voxel intensity value. By repeating this for all nearby Gaussians and summing up their contributions, we arrive at the voxel's final intensity value $c_{pixel}(x')$ as per Eq.~\ref{eq:colour-combine}. This is in contrast to traditional Gaussian Splatting for projection-based imaging~\cite{kerbl20233d}, where the 3D Gaussians are always projected onto the pinhole camera's 2D image plane and thus appear as 2D Gaussians; these 2D Gaussians can then only be used to reconstruct surfaces and not the interior of objects as our case requires.

\subsection{Initialisation of Gaussian positions}
\label{sec:gaussian_init}

3D Gaussian Splatting, due to its explicit-like representation nature, is well suited to generating novel views within the source field of view, but struggles outside it, since during training the Gaussians move and are optimised to over-fit the input/sample slices. Whilst this is often seen as a drawback in the natural image domain, it prevents hallucinations or artifacts – a much more critical consideration for medical applications. 

We choose to embrace this inherent limitation, and in doing so, design-in efficiency improvements. 

To ensure that a given number of Gaussians have the highest impact, we chose to initialise their means only within the volume bounded by the upper and lower source planes, rather than the volume encompassing all the source and target planes as in prior works~\cite{eid2025ultragauss} (see Supplementary Table \ref{tab:init-limits}). Accepting that renders of the sections of the blue planes (Fig. \ref{fig:3d-planes}) lying outside the yellow planes will always be of much worse quality than those within, we explicitly only initialise the Gaussians within the volume bounded by the yellow slices. We also make use of the fact that the source planes are rectangular in the $x-y$ plane, rather than assuming they are square, further reducing the initialisation volume. Furthermore, we found that random initialisation of Gaussian positions provided better results than uniform initialisation or a gaussian-weighted initialisation mask centered about the region of interest (see Supplementary Table \ref{tab:init-strategy}). It also removes the complexity of having to pre-reconstruct a low-quality volume from which to initialise the Gaussians~\cite{zha_r2-gaussian_2024}.

\begin{figure}[!t]
\centering % inkscapelatex=false,
\includegraphics[width=0.75\textwidth]{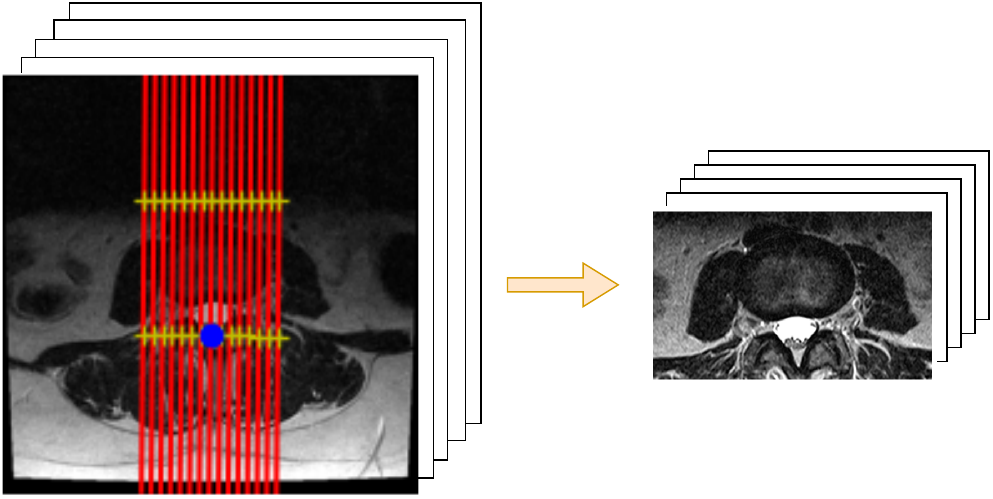}

\caption{\label{fig:crop} \textbf{Deriving a cropped axial region of interest:} Red lines indicate sagittal slices (coming in and out of the page, slicing from head to toe) that intersect the axial slice shown (see Fig.~\ref{fig:ex-paired} for example sagittal image). SpineNet is used to detect the intervertebral disc (IVD) pertaining to each spinal level in the sagittal scans. The intersection of the sagittal slices with the anterior (i.e. front side of the body, top of Fig.~\ref*{fig:crop}) and posterior (i.e. back side of the body, bottom of Fig.~\ref*{fig:crop}) boundaries of the detected IVD region are marked in yellow. The midpoint of the posterior intersection, indicated by the blue dot, is then used as a reference point to define the cropped region of interest in the axial scans. All axial slices that intersect the SpineNet-detected sagittal ROI are included to form the final axial IVD volume.} 
\end{figure}

We also crop the source images so that they are only slightly bigger than the region of interest (see Fig.~\ref{fig:crop}), ensuring that no Gaussians are wasted representing areas that would not then feature within the blue target slices (see Fig.~\ref{fig:3d-planes}). The width of the cropped axial image is chosen such that it is twice the lateral depth covered by the sagittal slices (i.e. horizontal distance covered by the red slices in Fig.~\ref{fig:crop}), and the height of the cropped axial image is chosen such that the aspect ratio of the extracted patch is 2:1, which is the dimension that the grading model expects ($H\times W=112 \times 224$).

To form 3D volumetric reconstructions quickly using Eqs.\ref{eq:mahalanobis-3d}-\ref{eq:colour-combine}, an efficient two-stage CUDA process ensures that Gaussians which are far from the image plane are first ignored, with each Gaussian then launching on its own GPU thread and only computing the Gaussian-pixel computations for pixels within its 95\% probability distribution ellipsoid, rather than all pixels on the image plane. This substantially reduces the number of times Eq. \ref{eq:mahalanobis-3d} is computed, and also means that the number of summations required in Eq. \ref{eq:colour-combine} is reduced too.

\subsection{Objectives, training and parameter tuning}
 
To learn and optimise the Gaussians' parameters, and thereby form a complete 3D reconstruction, we apply standard gradient descent and back-propagation to the photometric loss between the source ground truth (the originally non-aligned image) and the rendered source image (reconstruction of the non-aligned image) at the source slice pose. Specifically, our loss function is $\mathcal{L} = 0.7 \mathcal{L}_1 + 0.3(1-\text{SSIM})$. We found this to be a good balance between structural coherence (SSIM) and per-pixel accuracy ($\mathcal{L}_1$). It was also better than a mixture of our three image quality evaluation metrics – SSIM, PSNR and LPIPS. This is likely because PSNR is based off the L2 loss, which heavily penalises large errors (especially frequent at the start of training) thus often resulting in over-smoothing. Meanwhile, LPIPS often improved in opposite directions to SSIM and PSNR, making the overall optimisation worse as the gradient directions conflicted. 

\begin{figure}[!h]
\centering % inkscapelatex=false,
\includegraphics[width=0.9\textwidth]{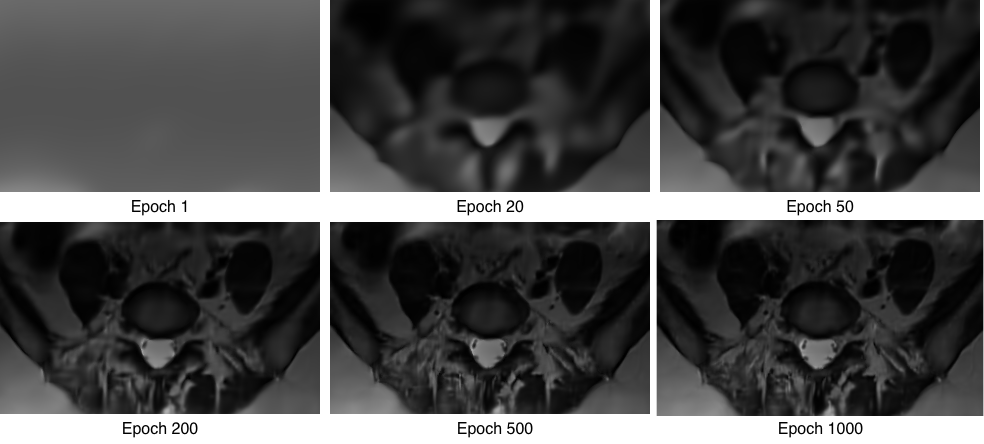}
\caption{\label{fig:ex-rendering} \textbf{3D Gaussian Splatting-reconstructed example across training epochs:} Reconstructed image across training epochs 1, 20, 50, 200, 500 and 1000. Parameters of the Gaussians are updated across epochs using a weighted sum of $\mathcal{L}_1$ and SSIM ($\mathcal{L} = 0.7 \mathcal{L}_1 + 0.3(1-\text{SSIM})$).} 
\end{figure}

To aid the optimisation process, we also use Gaussian resampling heuristics (Gaussian pruning, splitting and cloning)~\cite{kerbl20233d}. Once optimisation is complete, novel MR views at the target slices are then rendered by inputting the target slice poses. Fig.~\ref{fig:ex-rendering} shows the reconstruction of a novel view target at various training epochs.

\paragraph{Implementation and training details.}
We initialise 100,000 Gaussians, each with a colour $c_i = 0.5$ and alpha $\alpha_i = 0.623$. Exponentially decaying learning rate schedulers are applied for all Gaussian parameters. These start at $lr =\{0.00075,\\ 0.1, 0.01, 0.025\}$ and end at $lr = \{0.000075, 0.001, 0.001, 0.0025\}$ for $\mu_{i}, \Sigma_{i}, c_i,$ and $\alpha_i$ respectively. From iteration 4,000, every 7,500 iterations we grow any Gaussians with a Euclidean (L2) norm positional gradient above 0.01, as they are deemed to represent an area of high importance. Of these, if their largest standard deviation is $>0.1$, the Gaussian is split into two smaller Gaussians (by a factor of 1.6); otherwise, it is cloned into two identical Gaussians. Additionally, any Gaussians with an opacity $<0.4$ are deemed to not be contributing significantly, and so are removed. All hyper-parameter tuning was completed using scans from only eleven patients. These were then witheld from the test-set, as well as from the stenosis grading model training set (see Section \ref{sec:gradings}). As a result of the efficient initialisation discussed in Section \ref{sec:gaussian_init}, we found that we only need 100,000 Gaussians rather than 300,000 to achieve comparable results as previous works found for their applications~\cite{eid2025ultragauss}. Initialising more than 100,000 Gaussians increased the time taken to form the 3D reconstruction, with negligible improvements in image quality (see Supplementary Table \ref{tab:num-gaussians}). 

\section{Predicting radiological gradings using resampled scans}
\label{sec:gradings}

\begin{figure}[!h]
\centering 
\includegraphics[width=0.6\textwidth]{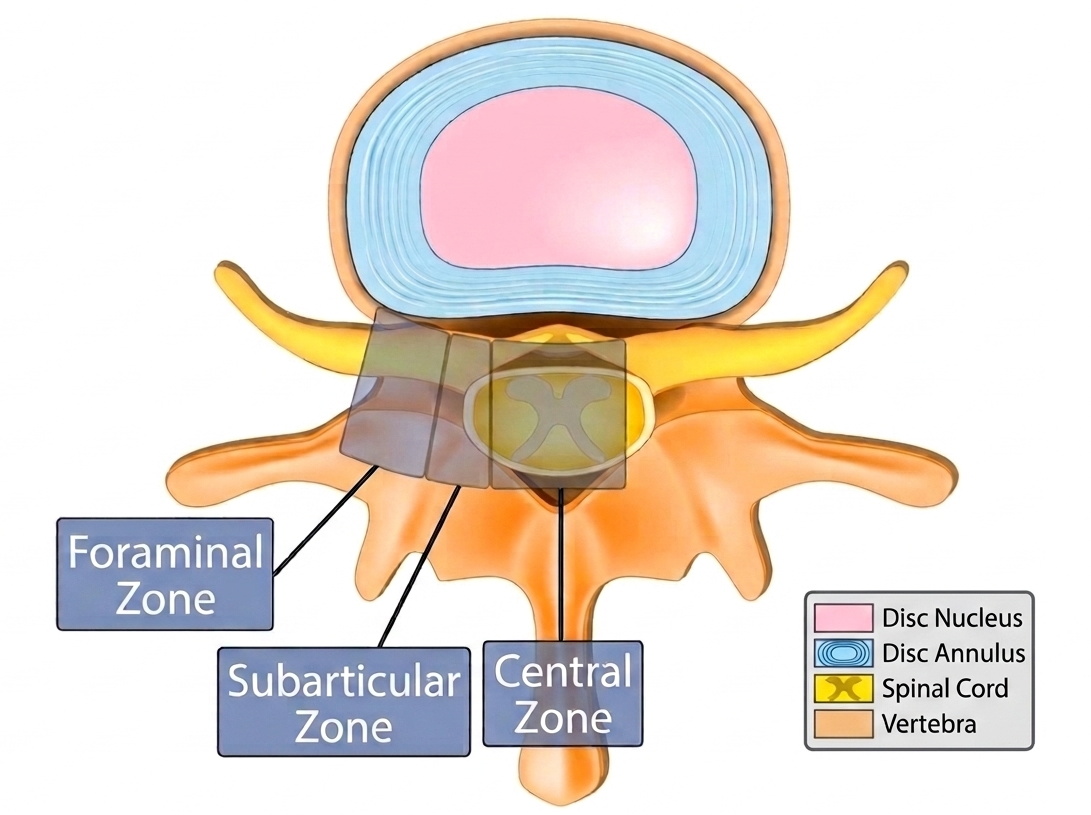}
\caption{\label{fig:stenosis-regions} \textbf{Stenosis regions:} The extrusion of the disc into each demarcated zone leads to compression of nerves and vascular structures in that region, causing stenosis. Image adapted from the Miami Neuroscience Center~\cite{miamnc_spinal_stenosis} using Nano Banana 2~\cite{nanobanana2026}.} 
\end{figure}

Stenosis is a degenerative condition that causes narrowing of the neurovascular structures around the spine~\cite{sobanski2023presentation}. It can cause back pain and disability in patients and can occur in different regions in the posterior of the disc~\cite{sobanski2023presentation, melancia2014spinal}. Fig.~\ref{fig:stenosis-regions} shows the different zones where localised  stenosis conditions can occur, relative to the disc and spinal canal.

\begin{figure}[!h]
\centering
\includegraphics[width=0.9\textwidth]{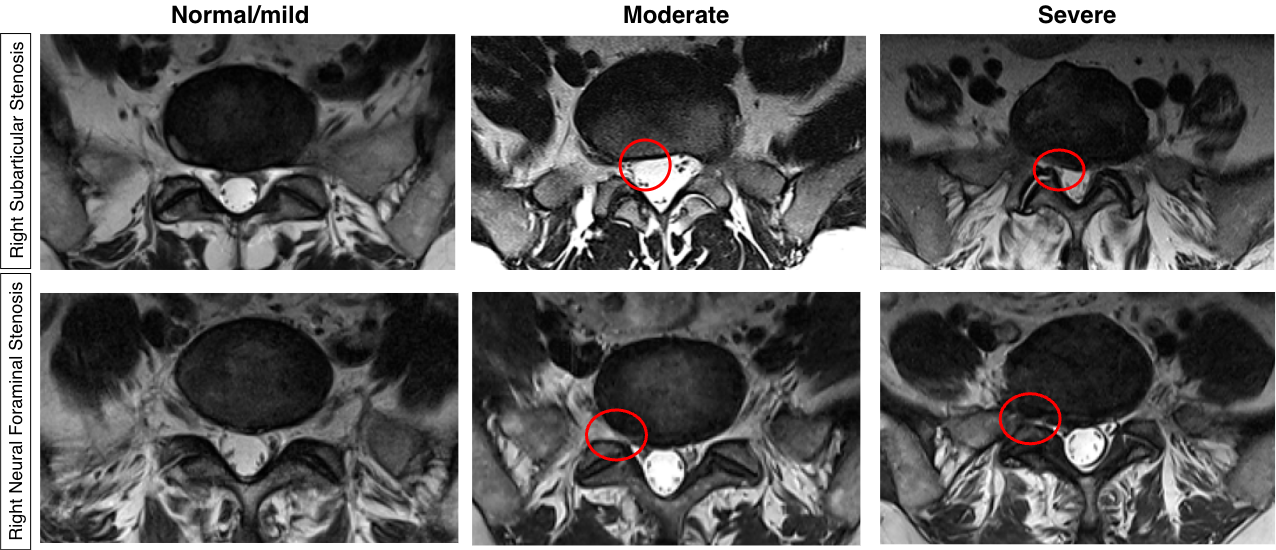}
\caption{\label{fig:ex-severities} \textbf{Stenosis severity gradings:} Examples of right subarticular and neural foraminal stenosis gradings. In the RSNA dataset, normal and mild are combined into a single category. The red ellipse indicates the location of the stenosis in moderate and severe images. The subarticular and foraminal zones are described in Fig.~\ref{fig:stenosis-regions}. Axial MRI is conventionally interpreted from the inferior-to-superior perspective.} 
\end{figure}

The presence and severity of degenerative spinal conditions are characterised by categorical grading scales ordered by severity. Fig.~\ref{fig:ex-severities} shows examples of neural foraminal stenosis and subarticular stenosis at varying degrees of severity. 

As a comparative approach to Gaussian Splatting, we also produce reconstructions using (1) \textbf{Voxel Interpolation (VI)}, which uses the inverse-distance weighted mean of nearest neighbour points of target plane coordinates for resampling and (2) \textbf{Cubic B-spline (CB)} resampling, which models a discrete signal as a piece-wise cubic function and samples at target points. More detailed descriptions of each method are provided in Section~\ref{sec:metrics-image-qual}. We compare L5-S1 stenosis grading performance over five test image types: (1) aligned (Test-\textbf{A}), (2) non-aligned (Test-\textbf{N}), (3) Voxel Interpolation resampled (Test-\textbf{R-VI}), (4) Cubic B-spline resampled (Test-\textbf{R-CB}) and (5) Gaussian Splatting resampled (Test-\textbf{R-GS}). For the Gaussian Splatting resampled set, we further consider models trained on three different training datasets: (1) aligned only (Train-\textbf{A}), (2) aligned + non-aligned (Train-\textbf{A\&N}), and (3) aligned + Gaussian Splatting resampled (Train-\textbf{A\&R-GS}). When training models 2 and 3, aligned scans were included alongside non-aligned or resampled scans due to the public dataset we use containing an insufficient number of non-aligned scans (and hence resampled scans too, since they are derived from non-aligned scans).

Each L5-S1 intervertebral disc (IVD) ROI was detected with Park \etal's~\cite{park2025multi} method, as shown in Fig.~\ref{fig:crop}. Train-\textbf{A} has 1,076 IVDs, Validation-\textbf{A} has 274 IVDs, Train-\textbf{N} has 200 IVDs and Validation-\textbf{N} has 52 IVDs. Resampled (\textbf{R}) images match the number of \textbf{N} scans across all splits, as they are reconstructed from non-aligned scans. The test set comprises of 119 patients who have both \textbf{A}ligned and \textbf{N}on-aligned scans (although in practice patients would only typically have \textbf{N}on-aligned scans) and are held-out from both train and validation sets. The train and validation sets were created using multi-label stratified sampling, ensuring no patient overlap and proportional label distribution.

\subsection{Grading model}

\begin{figure}[!h]
\centering
\includegraphics[width=0.9\textwidth]{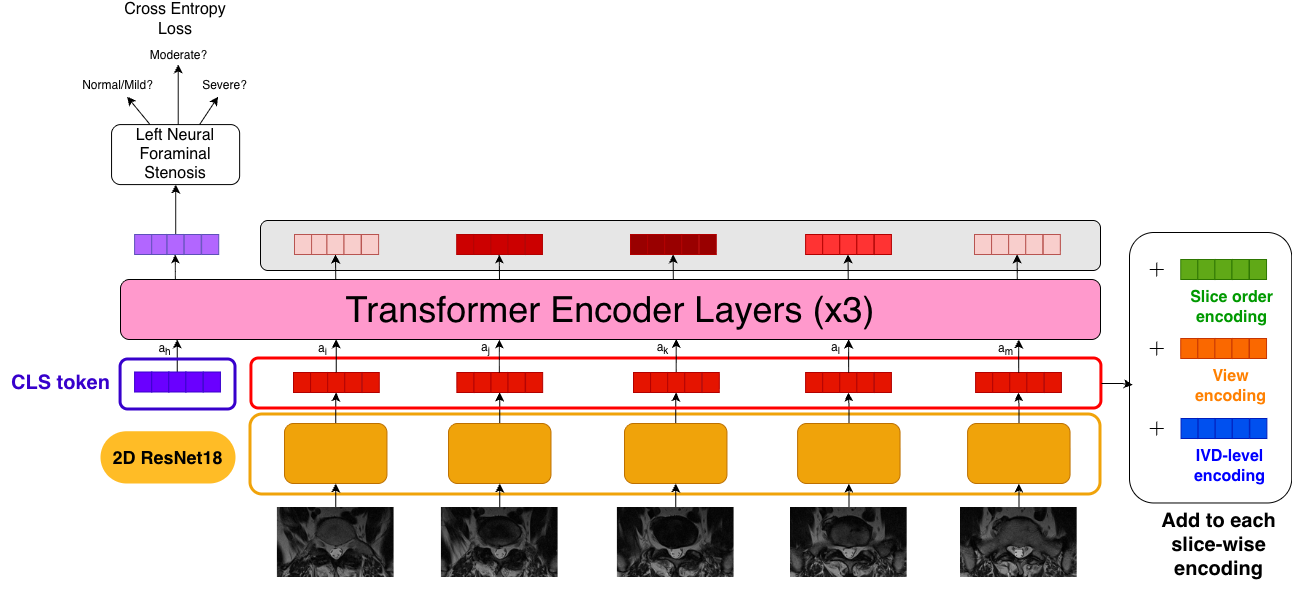}
\caption{\label{fig:grading-model} \textbf{Grading model:} We employ the same network described in~\cite{park2025multi}, where each slice in the volume is encoded using 2D Resnet18, then aggregated at volume-level using a global class (CLS) token. The output of the CLS token from the transformer encoder is passed to a linear layer that predicts the severity of each classification task.} 
\end{figure}

The grading model is based on the network of Park \etal~\cite{park2025multi} (see Fig.~\ref{fig:grading-model}). Each 2D slice in the volume is encoded using a shared 2D ResNet18. The slice features are aggregated using a global class (CLS) token prepended to the slice-wise encodings, then passed to a transformer encoder. The CLS output is fed to a linear head to predict condition severity. Training follows the specifications in~\cite{park2025multi}, with identical augmentations (random shift, intensity offset, flip, rotation, translation, scaling) for robustness. Separate models are trained for each direction (left/right) on each stenosis condition (neural foraminal and subarticular). The final model is selected based on the highest mean macro-average AUROC over 10 consecutive validation epochs. All model runs used a fixed random seed.

\section{Results}
 
We evaluate the reconstruction using (1) perceptual similarity and (2) performance on downstream grading tasks. As described in Section~\ref{sec:gradings}, the test set comprises of 119 patients with both aligned and non-aligned scans of the L5-S1 IVD. 

\subsection{Performance metrics}\label{sec:perf-metrics}
Learned Perceptual Image Patch Similarity (LPIPS using AlexNet weights) is our main image quality metric. LPIPS measures how similar two images look to the human eye via deep feature representations~\cite{zhang_unreasonable_2018}, which makes it more suitable than pixel-level metrics for radiology images presented to radiologists for clinical evaluation. 

Grading performance is measured by the macro-averaged Area Under the Receiver Operating Characteristic (AUROC) curves for each class. As the tasks are multi-class, class-wise One-vs-Rest (OvR) AUROC is computed and averaged for overall performance. We use AUROC instead of threshold-dependent metrics (e.g. balanced accuracy or F1) due to poor calibration across multi-class probabilities, which is exacerbated by data imbalance~\cite{guo2017calibration}. Test-time augmentation is applied for robustness by averaging softmax probabilities over 16 views per image including flips, shifts (4 pixels in each direction), rotations ($\pm5$\textdegree, $\pm10$\textdegree), scaling (95\%, 105\%), and intensity variations ($\pm0.02$ brightness, $\pm2\%$ contrast).

\begin{figure}[!h]
\centering
\includegraphics[width=0.9\textwidth]{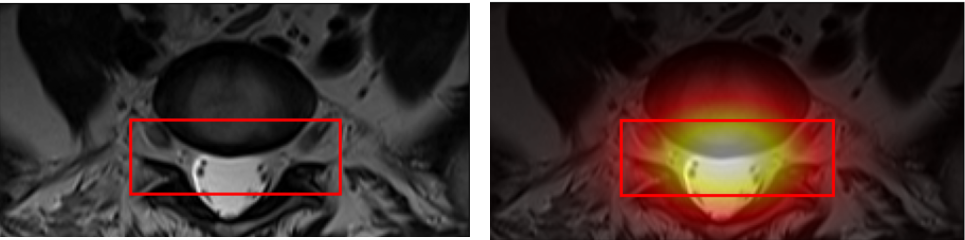}
\caption{\label{fig:gaussian-wgt} \textbf{Gaussian weight mask:} The red box shows the region relevant for stenosis grading. An anisotropic Gaussian weight mask focused on the region of interest attenuates the pixels outside of that region before computing LPIPS.} 
\end{figure}

\subsection{Reconstruction image quality}\label{sec:metrics-image-qual}

We compare Gaussian Splatting adapted for MRI with \textbf{Voxel Interpolation (VI)} and \textbf{Cubic B-spline (CB)} resampling methods. In \textbf{VI}, intensities from the non-aligned volume corresponding to nearest neighbours of each target volume coordinate (which can come from multiple planes of the source volume) are inverse distance-weighted and averaged to obtain the resampled intensities. In contrast, \textbf{CB} estimates intensity at target points by fitting a smooth chain of cubic segments that pass through or near source volume points. 

To ensure that the visual quality metrics focused on the clinically meaningful structures along the posterior of the disc, we computed the metrics using a Gaussian mask which provided spatially-varying weights (Fig.~\ref{fig:gaussian-wgt}). Before computing image quality metrics, the reconstructed images were translated by range of shifts up to $\pm20$ pixels relative to the target image. The shift maximizing PSNR was selected to account for any positional misalignment between acquisitions. All reported metrics were computed on this aligned pair.

\begin{figure}[h]
\centering
\includegraphics[width=0.9\textwidth]{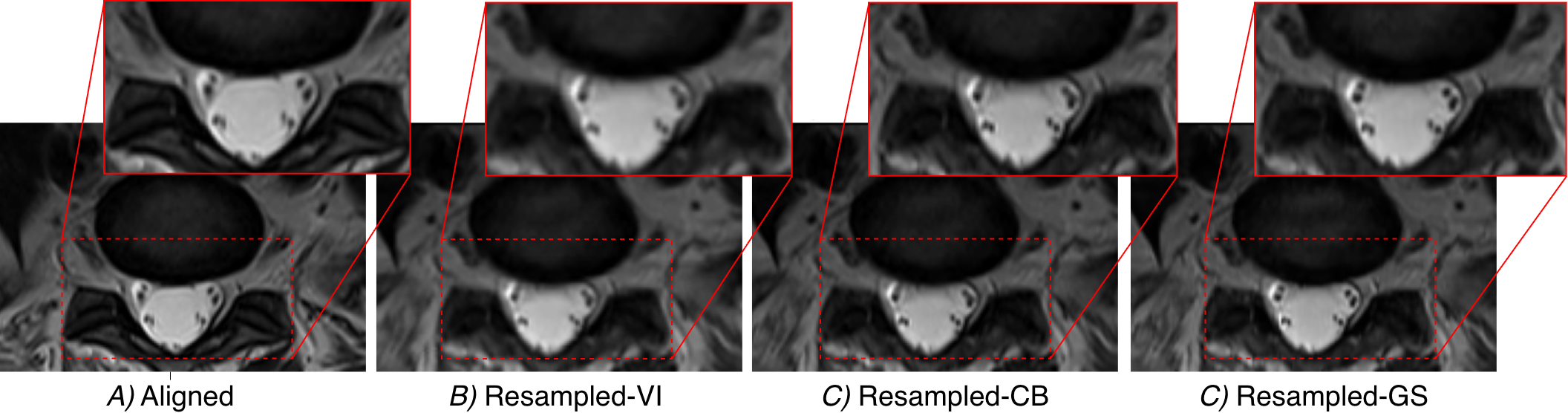}
\caption{\label{fig:ex-resample} \textbf{Resampled images:} \textit{A} shows the originally aligned slice. \textit{B} shows an image reconstructed using Voxel Interpolation (\textbf{R-VI}) and \textit{C} shows a slice reconstructed using Cubic B-spline resampling (\textbf{R-CB}). \textit{D} shows a slice reconstructed using Gaussian Splatting (\textbf{R-GS}). The red box shows the region relevant for grading.} 
\end{figure}

Gaussian Splatting achieves the best LPIPS (0.070) with an average of \textbf{13.58\%} improvement over the Voxel Interpolation method (0.081) and \textbf{2.78\%} improvement over the Cubic B-spline resampling (0.072). Supplementary Table \ref{tab:image-qual} shows pixel-wise image quality metrics for each resampling method alongside LPIPS. Fig.~\ref{fig:ex-resample} shows examples of slices resampled using each method compared with the originally aligned image for that patient. Supplementary Fig. \ref{fig:ex-3dgs-slices} shows a comparison of the full ground truth and GS-resampled volumes for a given patient.\\

% \clearpage
\subsection{Radiological grading performance}

If the reconstruction is successful, grading performance of the resampled scans (Test-\textbf{R}) should exceed grading performance of the non-aligned scans (Test-\textbf{N}), and ideally be close to the performance on natively aligned scans (Test-\textbf{A}). 

Results of GS are given in Table~\ref{tab:grad-perf}. As expected, Test-\textbf{A} performs the best across most conditions and training sets, but aligned images are not always available – not only in this dataset but also in standard clinical settings. Notably, Gaussian Splatting-resampled images (Test-\textbf{R-GS}) perform better than non-aligned images (Test-\textbf{N}), even when resampled views are not included in the training data of the grading model (Train-\textbf{A}/\textbf{A\&N}). Grading of Gaussian Splatting-resampled images (Test-\textbf{R-GS}) outperforms grading of non-aligned scans (Test-\textbf{N}) across all conditions and training data.

\begin{table}[h!]
\caption{\label{tab:grad-perf}\textbf{Macro average AUROC for grading tasks:} \textbf{A} indicates aligned scans, \textbf{N} non-aligned scans, and \textbf{R} resampled scans, with \textbf{GS} referring to Gaussian Splatting resampling. For each row, the training set of the grading model includes samples specified with a tick. Train-\textbf{A} includes 1,076 patients, and Train-\textbf{N}/\textbf{R-GS} include the same 200 patients. All test sets contain the same 119 patients. There are three classes for each condition: normal/mild, moderate and severe. The macro average AUROC is computed by averaging class-specific One-vs-Rest (OvR) AUROC. Red indicates suboptimal results on non-aligned images. Green indicates improvement from resampled scans. Grey indicates gold standard results on originally aligned scans. The best grading performance (per condition) from reconstructed images is bolded. Test-time augmentation is applied for robustness by averaging softmax probabilities over 16 views per image, as described in Section~\ref{sec:perf-metrics}.}

\centering
\setlength{\tabcolsep}{2.9pt}

\begin{tabular}{%
  l
  >{\centering\arraybackslash}m{0.94cm} >{\centering\arraybackslash}m{0.94cm} >{\centering\arraybackslash}m{0.94cm}              
  >{\centering\arraybackslash}m{0.94cm} >{\centering\arraybackslash}m{0.94cm} >{\centering\arraybackslash}m{0.94cm} >{\centering\arraybackslash}m{0.94cm}    
}
\toprule
\multirow{2}{*}[-2pt]{\textbf{Stenosis Condition}}
& \multicolumn{3}{c}{\textbf{Training Set}}
& \multicolumn{3}{c}{\textbf{Test Set}} \\
\cmidrule(lr){2-4} \cmidrule(lr){5-7}
& \textbf{A} & \textbf{N} & \textbf{R-GS}
& \textbf{N} & \textbf{R-GS}
& {\color[HTML]{9E9E9E}\textbf{A}} \\
\midrule
% ---------------- Neural Foraminal (L)
\multicolumn{1}{l}{\multirow{3}{*}{Neural Foraminal (L)}} &
\checkmark & & &
\cellcolor[HTML]{F4CCCC}0.731 &
\cellcolor[HTML]{D9EAD3}\textbf{0.815} &
{\color[HTML]{9E9E9E}0.832} \\
 & \checkmark & \checkmark & &
\cellcolor[HTML]{F4CCCC}0.727 &
\cellcolor[HTML]{D9EAD3}0.792 &
{\color[HTML]{9E9E9E}0.816} \\
 & \checkmark & & \checkmark &
\cellcolor[HTML]{F4CCCC}0.715 &
\cellcolor[HTML]{D9EAD3}0.812 &
{\color[HTML]{9E9E9E}0.838} \\
\midrule
% ---------------- Neural Foraminal (R)
\multicolumn{1}{l}{\multirow{3}{*}{Neural Foraminal (R)}} &
\checkmark & & &
\cellcolor[HTML]{F4CCCC}0.691 &
\cellcolor[HTML]{D9EAD3}0.779 &
{\color[HTML]{9E9E9E}0.819} \\
 & \checkmark & \checkmark & &
\cellcolor[HTML]{F4CCCC}0.736 &
\cellcolor[HTML]{D9EAD3}\textbf{0.812} &
{\color[HTML]{9E9E9E}0.807} \\
 & \checkmark & & \checkmark &
\cellcolor[HTML]{F4CCCC}0.766 &
\cellcolor[HTML]{D9EAD3}0.810 &
{\color[HTML]{9E9E9E}0.807} \\
\midrule
% ---------------- Subarticular (L)
\multicolumn{1}{l}{\multirow{3}{*}{Subarticular (L)}} &
\checkmark & & &
\cellcolor[HTML]{F4CCCC}0.695 &
\cellcolor[HTML]{D9EAD3}0.741 &
{\color[HTML]{9E9E9E}0.797} \\
 & \checkmark & \checkmark & &
\cellcolor[HTML]{F4CCCC}0.705 &
\cellcolor[HTML]{D9EAD3}\textbf{0.767} &
{\color[HTML]{9E9E9E}0.798} \\
 & \checkmark & & \checkmark &
\cellcolor[HTML]{F4CCCC}0.734 &
\cellcolor[HTML]{D9EAD3}0.766 &
{\color[HTML]{9E9E9E}0.824} \\
\midrule
% ---------------- Subarticular (R)
\multicolumn{1}{l}{\multirow{3}{*}{Subarticular (R)}} &
\checkmark & & &
\cellcolor[HTML]{F4CCCC}0.695 &
\cellcolor[HTML]{D9EAD3}0.743 &
{\color[HTML]{9E9E9E}0.800} \\
 & \checkmark & \checkmark & &
\cellcolor[HTML]{F4CCCC}0.706 &
\cellcolor[HTML]{D9EAD3}0.790 &
{\color[HTML]{9E9E9E}0.798} \\
 & \checkmark & & \checkmark &
\cellcolor[HTML]{F4CCCC}0.734 &
\cellcolor[HTML]{D9EAD3}\textbf{0.794} &
{\color[HTML]{9E9E9E}0.814} \\
\bottomrule
\end{tabular}
\end{table}

\begin{table}[h!]
\caption{\label{tab:base-compare}\textbf{Comparison of grading using resampled scans:} Macro average AUROC for grading model trained with natively aligned samples (Train-\textbf{A}: 1,076 patients) and evaluated on scans resampled using each reconstruction method (Test-\textbf{R}: 119 patients). There are three classes for each condition: normal/mild, moderate and severe. The macro average AUROC is computed by averaging class-specific One-vs-Rest (OvR) AUROC. Test-time augmentation is applied for robustness by averaging softmax probabilities over 16 views per image, as described in Section~\ref{sec:perf-metrics}. The best grading performance (per condition) is bolded.} 
\centering
\setlength{\tabcolsep}{5pt}  % set column separation
\begin{tabular}{cccc}
\toprule
Stenosis Condition & Voxel Interpolation & Cubic B-spline & Gaussian Splatting\\
\midrule
Neural Foraminal (L) & 0.783 & 0.790 & \textbf{0.815} \\
Neural Foraminal (R) & 0.770 & 0.773 & \textbf{0.779} \\
Subarticular (L) & 0.722 & 0.722 & \textbf{0.741}\\
Subarticular (R) & 0.735 & 0.740 & \textbf{0.743} \\
\bottomrule
\end{tabular}
\end{table}

\noindent \textbf{Comparison with baselines.} Table~\ref{tab:base-compare} shows that Gaussian Splatting-resampled scans outperform Voxel Interpolation and Cubic B-spline resampled images across all conditions, i.e.\ grading of Gaussian-Resampled images leads to better stenosis degree classification than grading of images resampled using Voxel Interpolation or Cubic B-spline resampling.

\section{Discussion of results}
 
GS-resampled scans (Test-\textbf{R-GS}) not only achieve better LPIPS image quality metrics, but also outperform Voxel Interpolation and Cubic B-spline-resampled scans for \textit{all grading scenarios}, demonstrating that 3D GS resampling can effectively reconstruct misaligned MRI scans and improve downstream grading accuracy. Compared to using non-aligned MRI scans that suboptimally intersect the relevant anatomy, which in practice make up a large number of spinal acquisitions, our MRI Gaussian Splatting reconstruction method achieves a 9.1\% improvement in AUROC, providing a potential route to more accurate and efficient patient diagnostics without needing another resource-intensive MRI scan. 

One limitation is the limited number of non-aligned (\textbf{N}) samples. A useful experiment would be to train solely on \textbf{N} samples and evaluate on resampled (\textbf{R}) images, but there were too few \textbf{N} samples to train a useful grading model. Future work includes experiments on other MR views (e.g.\ sagittal) and modalities (e.g.\ T1w, STIR) for a wider range of degenerative disc conditions, or diseases involving other anatomical structures, such as diffuse cancer in vertebral bodies.

\section{Conclusion}

3D Gaussian Splatting can be used to render novel views of MRI that show a clearer view of the relevant anatomy, leading to better automated grading performance compared to anatomically non-aligned MR acquisitions. This result holds whether or not the grading model's training data include reconstructed samples alongside originally aligned scans. While our experiments show the efficacy of our reconstruction method on spinal stenosis, this method can be applied to any non-projection-based medical imaging modality and target anatomy. Our code is available at: \url{https://www.robots.ox.ac.uk/~vgg/research/MRIGauss/}.

% \clearpage  % TODO FINAL: This \clearpage needs to be removed from both review and camera-ready versions.

\section*{Acknowledgments.} We are grateful to our funders: EPSRC CDT in Autonomous Intelligent Machines and Systems (EP/S024050/1), EPSRC CDT in Health Data Science (EP/S02428X/1), EPSRC programme grant Visual AI (EP/T025872/1), Royal Academy of Engineering (RF/201819/ 18/163), Amazon Web Services, and Bill \& Melinda Gates Foundation.

\section*{Disclosure of interests.} The authors have no competing interests to declare that are relevant to the content of this article.

\newpage
\bibliographystyle{splncs04}
\bibliography{main_noDOI}

\clearpage
\setcounter{table}{0}      % Reset table numbers to 0
\setcounter{figure}{0}     % Reset figure numbers to 0

\title{SUPPLEMENTARY MATERIAL:\\Rendering Novel Views of MRI Using 3D Gaussian Splatting} 
\titlerunning{ }
\author{}
\authorrunning{}
\institute{}
\maketitle

\begin{table}[h!]
\caption{\label{tab:init-limits}\textbf{Effect of the choice of volume in which gaussians are initialised} on reconstruction quality. For each metric, resampled images were compared against each originally-aligned image and weighted by a Gaussian mask to focus on the relevant region for grading. Each metric was averaged across slices in a volume to obtain a volume average, then further averaged across the entire hyper-parmameter tuning set (11 patients). Best scores are bolded.}
\centering
\setlength{\tabcolsep}{5pt}  % set column separation
\renewcommand{\arraystretch}{1.425}
\begin{tabular}{>{\raggedright\arraybackslash}m{7cm}
                >{\centering\arraybackslash}m{1cm}
                >{\centering\arraybackslash}m{1cm}
                >{\centering\arraybackslash}m{1cm}}
\toprule
Init. Limits & PSNR\,\textuparrow & SSIM\,\textuparrow & LPIPS\,\textdownarrow \\
\midrule
Randomly within the $z$ limits of the cuboid bounding the \textit{source} slices & \textbf{23.013} & \textbf{0.600} & \textbf{0.057} \\
Randomly within the $z$ limits of the cuboid bounding the \textit{source and target} slices & 22.735 & 0.595 & 0.059 \\
Randomly within a \textit{unit cube} & 22.496 & 0.593 & 0.066 \\
\bottomrule
\end{tabular}
\end{table}

\begin{table}[h!]
\caption{\label{tab:init-strategy}\textbf{Effect of initialisation strategy} on reconstruction quality. Metrics were computed as in Table~\ref{tab:init-limits}.}
\centering
\setlength{\tabcolsep}{5pt}  % set column separation
\renewcommand{\arraystretch}{1.425}
\begin{tabular}{>{\raggedright\arraybackslash}m{7cm}
                >{\centering\arraybackslash}m{1cm}
                >{\centering\arraybackslash}m{1cm}
                >{\centering\arraybackslash}m{1cm}}
\toprule
Init. Strategy & PSNR\,\textuparrow & SSIM\,\textuparrow & LPIPS\,\textdownarrow \\
\midrule
\textit{Randomly} within the $z$ limits of the cuboid bounding the source slices & \textbf{23.013} & \textbf{0.600} & \textbf{0.057} \\
\textit{Uniformly} within the $z$ limits of the cuboid bounding the source slices & 22.913 & 0.598 & 0.058 \\
\textit{Gaussian-weighted initialisation} within the $x$-$y$ plane, randomly within the $z$ limits of the cuboid bounding the source slices & 23.000 & 0.588 & 0.060 \\
\bottomrule
\end{tabular}
\end{table}

\begin{table}[h!]
\caption{\label{tab:num-gaussians}\textbf{Effect of number of Gaussians} on reconstruction quality. Metrics were computed as in Table~\ref{tab:init-limits}.}
\centering
\setlength{\tabcolsep}{5pt}  % set column separation
\begin{tabular}{cccc}
\toprule
No. of Gaussians & PSNR\,\textuparrow & SSIM\,\textuparrow & LPIPS\,\textdownarrow \\
\midrule
25,000    & 22.575 & 0.589 & 0.063 \\
50,000    & 22.828 & 0.590 & 0.060 \\
100,000   & 23.013 & 0.600 & \textbf{0.057} \\
500,000   & 23.173 & 0.609 & \textbf{0.057} \\
1,000,000 & \textbf{23.255} & \textbf{0.613} & \textbf{0.057} \\
\bottomrule
\end{tabular}
\end{table}

\begin{table}[h!]
\caption{\label{tab:image-qual}\textbf{Image quality metrics} for Voxel Interpolation (VI) and Gaussian Splatting (GS) resampling. For each metric, resampled images were compared against each originally-aligned image and weighted by a Gaussian mask to focus on the relevant region for grading. Each metric was averaged across slices in a volume to obtain a volume average, then further averaged across the entire test set (119 patients).} 
\centering
% \fontsize{8pt}{10pt}\selectfont
\setlength{\tabcolsep}{5pt}  % set column separation
\begin{tabular}{cccc}
\toprule
Method & PSNR\,\textuparrow & SSIM\,\textuparrow & LPIPS\,\textdownarrow \\
\midrule
Voxel Interpolation & \textbf{23.080} & \textbf{0.609} & 0.081 \\
Cubic B-spline & 22.865 & 0.594 & 0.072 \\
Gaussian Splatting & 22.500 & 0.571 & \textbf{0.070} \\
% \addlinespace[2pt]
\bottomrule
\end{tabular}
\end{table}

\begin{figure}[h!]
\centering
\includegraphics[width=0.76\textwidth]{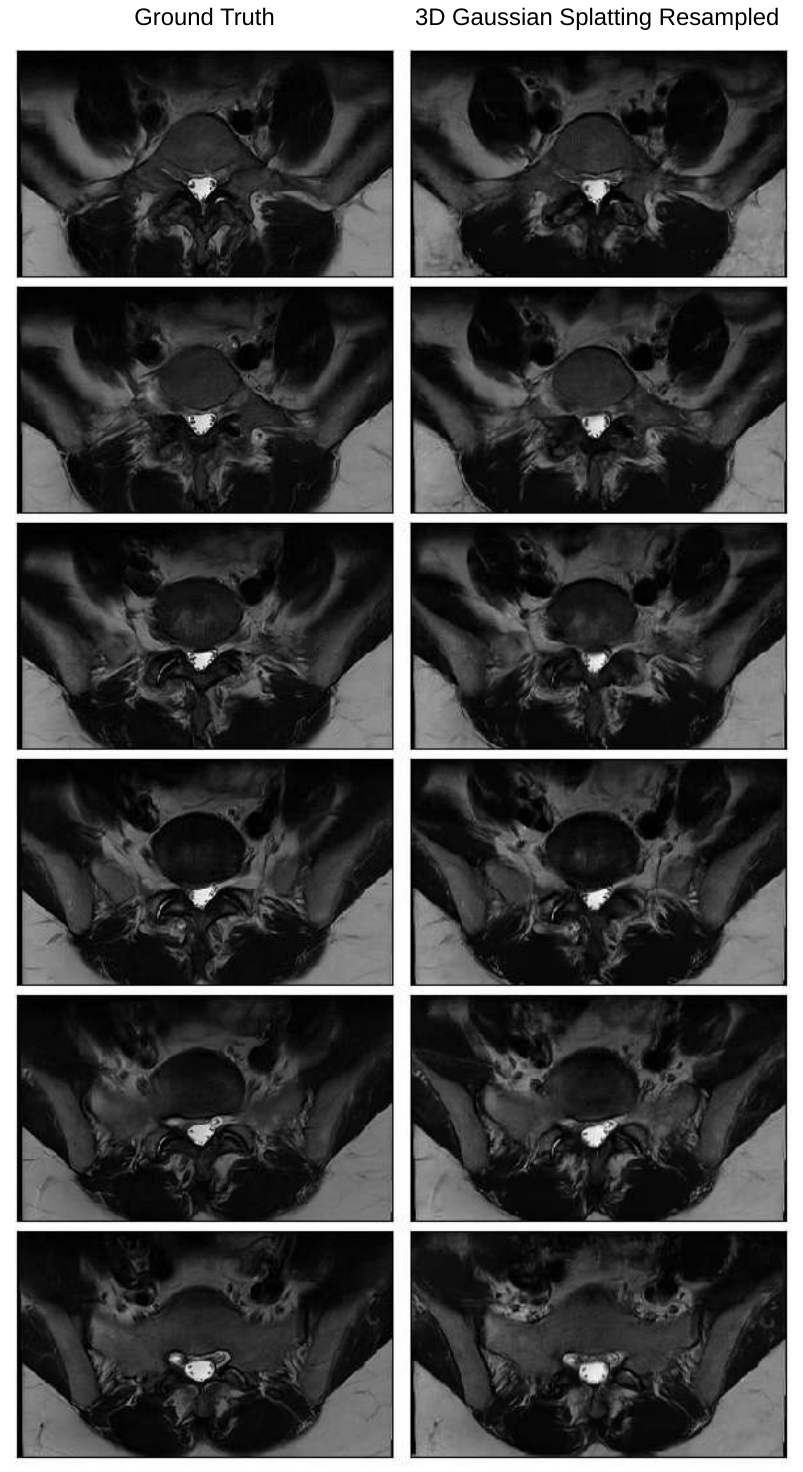}
\caption{\label{fig:ex-3dgs-slices} \textbf{Examples of rendered slices:} Ground truth images acquired from an aligned-view volume compared against views rendered from a 3D Gaussian Splatting reconstruction from a non-aligned source volume.} 
\end{figure}

\end{document}